\makeatletter
\def\cl@chapter{}
\makeatother

\documentclass[twocolumn,natbib,english]{svjour3}        

\makeatletter
\if@twocolumn
  \renewcommand\normalsize{\@setfontsize\normalsize\@xpt{12.5pt}\abovedisplayskip=3 mm plus6pt minus 4pt
    \belowdisplayskip=3 mm plus6pt minus 4pt
    \abovedisplayshortskip=0.0 mm plus6pt
    \belowdisplayshortskip=2 mm plus4pt minus 4pt
    \let\@listi\@listI}

  \renewcommand\small{\@setfontsize\small{8.5pt}\@xpt
    \abovedisplayskip 8.5\p@ \@plus3\p@ \@minus4\p@
    \abovedisplayshortskip \z@ \@plus2\p@
    \belowdisplayshortskip 4\p@ \@plus2\p@ \@minus2\p@
    \def\@listi{\leftmargin\leftmargini
      \parsep 0\p@ \@plus1\p@ \@minus\p@
      \topsep 4\p@ \@plus2\p@ \@minus4\p@
      \itemsep0\p@}\belowdisplayskip \abovedisplayskip}
\else
  \if@smallext
    \renewcommand\normalsize{\@setfontsize\normalsize\@xpt\@xiipt
      \abovedisplayskip=3 mm plus6pt minus 4pt
      \belowdisplayskip=3 mm plus6pt minus 4pt
      \abovedisplayshortskip=0.0 mm plus6pt
      \belowdisplayshortskip=2 mm plus4pt minus 4pt
      \let\@listi\@listI}

    \renewcommand\small{\@setfontsize\small\@viiipt{9.5pt}\abovedisplayskip 8.5\p@ \@plus3\p@ \@minus4\p@
      \abovedisplayshortskip \z@ \@plus2\p@
      \belowdisplayshortskip 4\p@ \@plus2\p@ \@minus2\p@
      \def\@listi{\leftmargin\leftmargini
        \parsep 0\p@ \@plus1\p@ \@minus\p@
        \topsep 4\p@ \@plus2\p@ \@minus4\p@
        \itemsep0\p@}\belowdisplayskip \abovedisplayskip}
  \else
    \renewcommand\normalsize{\@setfontsize\normalsize{9.5pt}{11.5pt}\abovedisplayskip=3 mm plus6pt minus 4pt
      \belowdisplayskip=3 mm plus6pt minus 4pt
      \abovedisplayshortskip=0.0 mm plus6pt
      \belowdisplayshortskip=2 mm plus4pt minus 4pt
      \let\@listi\@listI}

    \renewcommand\small{\@setfontsize\small\@viiipt{9.25pt}\abovedisplayskip 8.5\p@ \@plus3\p@ \@minus4\p@
      \abovedisplayshortskip \z@ \@plus2\p@
      \belowdisplayshortskip 4\p@ \@plus2\p@ \@minus2\p@
      \def\@listi{\leftmargin\leftmargini
        \parsep 0\p@ \@plus1\p@ \@minus\p@
        \topsep 4\p@ \@plus2\p@ \@minus4\p@
        \itemsep0\p@}\belowdisplayskip \abovedisplayskip}
  \fi
\fi

\makeatother

\usepackage{cmap}

\RequirePackage[shrink=40,
  final,expansion=alltext,protrusion=alltext-nott]{microtype}\DisableLigatures{encoding = T1, family = tt* }

\usepackage[ngerman,main=english]{babel}
\addto\extrasenglish{\languageshorthands{ngerman}\useshorthands{"}}

\usepackage{ifluatex}
\ifluatex
  \usepackage{fontspec}
  \usepackage[english]{selnolig}
\fi

\ifluatex
 \else
\usepackage[rm={oldstyle=false,proportional=true},sf={oldstyle=false,proportional=true},tt={oldstyle=false,proportional=true,variable=false},qt=false]{cfr-lm}
\fi

\ifluatex
\else
\usepackage[T1]{fontenc}
  \usepackage[utf8]{inputenc} \fi

\smartqed  

\usepackage{graphicx}

\usepackage{upquote}

\usepackage{booktabs}

\usepackage{paralist}

\usepackage{csquotes}

\usepackage{textcmds}

\usepackage{url}
\makeatletter
\g@addto@macro{\UrlBreaks}{\UrlOrds}
\makeatother

\usepackage{xcolor}

\usepackage{listings}
\renewcommand{\lstlistingname}{List.}
\usepackage{chngcntr}
\AtBeginDocument{\counterwithout{lstlisting}{section}}

\usepackage{pdfcomment}

\newcommand*{\doi}[1]{\href{https://doi.org/\detokenize{#1}}{DOI: \detokenize{#1}}}

\usepackage{stfloats}
\fnbelowfloat

\usepackage[all]{hypcap}

\usepackage[acronym,indexonlyfirst,nomain]{glossaries}
\glsdisablehyper

\usepackage[capitalise,nameinlink]{cleveref}
\crefname{section}{Sect.}{Sect.}
\Crefname{section}{Section}{Sections}
\crefname{listing}{\lstlistingname}{\lstlistingname}
\Crefname{listing}{Listing}{Listings}

\usepackage{xspace}

\DeclareFontFamily{U}{MnSymbolC}{}
\DeclareSymbolFont{MnSyC}{U}{MnSymbolC}{m}{n}
\DeclareFontShape{U}{MnSymbolC}{m}{n}{
  <-6>    MnSymbolC5
  <6-7>   MnSymbolC6
  <7-8>   MnSymbolC7
  <8-9>   MnSymbolC8
  <9-10>  MnSymbolC9
  <10-12> MnSymbolC10
  <12->   MnSymbolC12}{}
\DeclareMathSymbol{\powerset}{\mathord}{MnSyC}{180}

\hypersetup{colorlinks=true,
  linkcolor=black,citecolor=black,urlcolor={blue!80!black},
  pdfstartview=Fit,
  breaklinks=true}

\journalname{Journal}

\usepackage[math]{blindtext}
\usepackage{mwe}
 \usepackage[type={CC},modifier={by},version={4.0},imagewidth=5em]{doclicense}
\usepackage{stmaryrd}
\usepackage{subcaption}
\usepackage[frozencache=true,cachedir=_minted-paper]{minted}
\usepackage{tikz}
\usetikzlibrary{arrows,shapes,positioning,fit,calc,decorations.pathreplacing,shapes.misc,shapes.geometric,decorations.pathmorphing,tikzmark,chains,tikzmark,scopes,matrix,backgrounds,patterns,overlay-beamer-styles,shadows,graphs}
\usepackage{booktabs}
\usepackage{array}
\newcolumntype{H}{>{\setbox0=\hbox\bgroup}c<{\egroup}@{}}
\newcolumntype{P}{>{(}r<{)}}

\usepackage{longtable}
\usepackage{multirow}
\usepackage{diagbox}

\begin{document}

\title{Easing Maintenance of Academic Static Analyzers}

\author{Raphaël Monat \and Abdelraouf Ouadjaout \and Antoine Miné}

\authorrunning{Monat \and Ouadjaout \and Miné}

\institute{
  Univ. Lille, Inria, CNRS, Centrale Lille, UMR 9189 CRIStAL, F-59000 Lille, France
  \and \\LIP6, Sorbonne Université, F-75005, Paris, France
  \and \\LIP6, Sorbonne Université, F-75005, Paris, France
}
\date{Revised: 05/11/2024}

\maketitle

\begin{abstract}
  Academic research in static analysis produces software implementations.
  These implementations are time-consuming to develop and some need to be maintained in order to enable building further research upon the implementation.
  While necessary, these processes can be quickly challenging.
  This article documents the tools and techniques we have come up with to simplify the maintenance of Mopsa since 2017.
  Mopsa is a static analysis platform that aims at being sound.
  First, we describe an automated way to measure precision that does not require any baseline of true bugs obtained by manually inspecting the results.
  Further, it improves transparency of the analysis, and helps discovering regressions during continuous integration.
  Second, we have taken inspiration from standard tools observing the concrete execution of a program to design custom tools observing the abstract execution of the analyzed program itself, such as abstract debuggers and profilers. Finally, we report on some cases of automated testcase reduction.

\end{abstract}

\section{Introduction}

One of the products of academic research in static analysis is the implementation of tools, which are used to illustrate and evaluate approaches.
During their lifetime, these tools have to be maintained -- in particular to enable further research.
This research  can be performed by the same authors, new members of the same group or by other groups wanting to build upon this tool.
While necessary, debugging and maintenance of static analyzers can quickly turn out to be time-consuming, as static analyzers perform highly technical reasoning.
In addition, this maintenance is not the main purpose of academic jobs -- and similarly, industrial developers will be looking to reduce their maintenance costs as much as possible.
Due to its nature, the maintenance problem is common to all researchers in the community.
Some practices are considered as folklore, which might explain why practices developed after years of experience are not systematically documented nor shared between groups.
In this paper, we document the tools and techniques we use to simplify our maintenance of Mopsa, the static analysis platform we have been developing since 2017.
We hope this article, following the precursor work of \cite{DBLP:conf/pldi/AndreasenMN17}, will have a twofold impact, by being useful to other researchers (and especially newcomers such as students), and by motivating other groups to share their practices. 

\paragraph{Contributions.}
\begin{itemize} 
\item We describe a novel way of reporting analyses results, and measuring precision based on selectivity (\Cref{sec:selectivity}).
  It is automatic, enhances transparency of the analysis, and takes into account the complexity of the expressions in the analyzed program (and not merely the program size).
  We can then leverage analyses results to detect soundness and precision regressions, thanks to a set of real-world, open-source benchmarks used in our continuous integration.
\item In \Cref{sec:hooks}, we show how plug-in observers to the analysis can help in the development of coverage and profiling tools working on the abstract interpretation of the program, while standard tools provide different results by working on the concrete execution of the analyzer.
\item We showcase an abstract debugger interface allowing interactive exploration of the abstract interpretation of the analyzed program, facilitating use and debugging (\Cref{sec:interactive}). This abstract debugger is available both as a command-line interface and in IDEs supporting the Debug Adapter Protocol.
\item We report on our experience in applying automated testcase reduction tools (initially used on compilers) on static analyzers (\Cref{sec:creduce}). These tools ease the process of isolating, analyzing and fixing precision and soundness issues in the analyzer.
  Indeed, this process is essentially manual and time-consuming, and bugs are often initially identified on example runs too large to inspect manually.
  Collaboration between Mopsa and automated testcase reduction is a two-way street: Mopsa enables seamless testcase reduction for multi-file projects with complex commands such as \texttt{GNU coreutils}.
  This process is enabled thanks to the \texttt{mopsa-build} utility instrumenting the compilation process, which can then be leveraged to output a single preprocessed file usable by testcase reduction tools. The generated file corresponds to a kind of AST-level linking.
\end{itemize}

In particular, \Cref{sec:hooks,sec:interactive} highlight a systematic connection between standard tools observing the \emph{concrete} execution of the \emph{abstract interpreter} and custom tools (abstract debuggers, profilers) we developed, which observe the \emph{abstract} execution of the \emph{analyzed} program itself.

\section{An Overview of Mopsa}

Mopsa (\textit{Modular Open Platform for Static Analysis}) is a publicly-available (\cite{mopsa-soft}) and open-source framework for the development of static analyzers based on the theory of abstract interpretation by \cite{CC77}.
\cite{mopsa,DBLP:phd/hal/Monat21} describe its flexible and modular architecture that makes it extensible in many aspects.
Mopsa aims at simplifying the exploration of new ideas and the development of static analyzers, while providing mature implementations for selected mainstream languages.
Mopsa participated in 2023 and 2024 to the Software-Verification Competition (SV-Comp), following contributions from \cite{DBLP:conf/tacas/MonatOM23,DBLP:conf/tacas/MonatMPBOM24}.
In 2024, Mopsa obtained the first place in the SoftwareSystems track of SV-Comp.

\subsection{Languages}
Mopsa supports the analysis of multiple languages, such as C (\cite{DBLP:conf/sas/JournaultMO18,mopsa-cstubs}), Python (\cite{PythonTypes,PythonSensitivities}), combination of Python and C (\cite{MonatOM21}) and Michelson (\cite{michelson-guillaume}) languages.
Unlike most multi-language static analyzers, Mopsa does not rely on translating programs to a fixed intermediate representation before starting the analysis.
These intermediate representations are generally very low-level (e.g., three-address code, stack machine, small C subsets).
While intermediate representations simplify the reuse of abstractions among different languages, they may suffer from information loss -- such as high-level control-flow structures and data structures -- during the initial syntactic translation.
For example, LLVM forgets whether integer types are signed or unsigned, while transformation to 3-address code puts a strain on relational domains to maintain precision, as shown by \cite{namjoshi18}.
In addition, a fixed intermediate representation may not support all kinds of languages and paradigms (e.g, dynamically typed, object oriented programming languages, functional languages), limiting the generality of the framework.

On the contrary, Mopsa has an extensible AST (\textit{Abstract Syntax Tree}) that represents the union of the original ASTs of the supported languages.
This approach preserves the original semantics of the program, allowing developers to reason on it directly.
However, Mopsa is not a disjoint union of separate analyzers, but tries to reuse abstractions between different languages.
This is done via two key mechanisms: \textit{semantic rewriting} and \textit{delegation}.

Instead of an initial syntactic translation, as performed by classic analyzer frontends, Mopsa can rewrite a statement to other languages \textit{during the analysis}.
This dynamic translation can exploit the inferred values of variables to produce more precise, or more efficient transformations.
For example, an integer addition in C can be translated to a mathematical addition if the sum of the operands fits within the range of the expression type.
Similarly, an addition in Python can be translated to a mathematical addition if the operands have only values of integer types, and do not implement the special methods \verb|__add__| and \verb|__radd__|.
Mopsa provides ready-to-use abstractions for mathematical integers, such as intervals and polyhedra.
Therefore, both C and Python analyses can \textit{delegate} the processing of the translated expression to one of these domains, which enables abstraction reuse between different analyses.

\subsection{Precision}
\label{sec:sub:precision}

Sound static analyzers may be imprecise due to coarse over-approximations.
This problem can be overcome by designing more accurate abstractions.
These improvements are generally very localized, targeting specific language constructs, such as how integer values are encoded or how loops are iterated.
Therefore, it is important to decompose the global abstraction into smaller pieces, called \textit{domains}, that are responsible for different parts of the language.
Domains handling the same statements or expressions should be easily exchangeable without impacting the remaining analysis, or seamlessly combined in a reduced product to enhance precision.

In Mopsa, the definition of the global abstraction is done with a JSON file called a \textit{configuration}, that lists the domains to include in the analysis as well as their order of execution and relations; and domains implement a common unified OCaml signature.

\paragraph{Unified Signature.}
From a developer perspective, the unified signature simplifies the integration of new domains.
This signature provides an API to allow cooperation between domains while preserving low coupling among them.
For example, when delegating the execution of a statement, a domain cannot call another domain directly.
Instead, it asks the framework to find within the configuration the appropriate domain(s) implementing the transfer function of the statement.

\paragraph{Configurations.}
From a user perspective, the JSON format of configurations gives a simplified way to define a new analysis by combining existing domains.
Mopsa supports different types of \textit{combinators} to compose domains, such as sequences and reduced products.
Many ready-to-use configurations with various tradeoffs between precision and efficiency are also provided.

\subsection{Properties}
Similarly to most static analyzers, Mopsa can verify classic reachability properties, such as absence of runtime errors in C and uncaught exceptions in Python.
In addition, some analyses are experimented in Mopsa, for in-progress research projects.
In particular, some target more complex reachability properties.
\cite{exploitability-francesco} verify the user-exploitability of alarms in C codes.
\cite{endianness-david} developed ways to prove portability of C programs between architectures with different endianness.
\cite{DBLP:conf/esop/MonatFM24} have developed an analysis targeting the rounding-sensitivity of date computations, in the context of legal implementations using the Catala programming language (\cite{merigoux2021catala}).

These kinds of reachability properties rely on computing necessary post-conditions with an over-approximating forward analysis.
The aim of these kinds of analysis is certifying the correctness of the program.
Mopsa was recently extended to support the computation of sufficient preconditions via an under-approximating backward analysis described by \cite{witness-marco}.
This kind of analysis is able to generate counter-examples for certifying program incorrectness.

\section{Measuring Precision}
\label{sec:selectivity}

The precision of static analyzers is a cornerstone of experimental evaluations developed to evaluate the benefits of new approaches.
It can also be leveraged to detect changes and regressions during tool development.
We provide a quick survey of classic approaches to measuring precision in static analysis,  before introducing a new way to automatically compute precision, which improves the transparency of the analysis and which can be numerically quantified.
Then, we highlight how our approach can be naturally leveraged to compare analysis results, and how this comparison is used to detect changes during development, through continuous integration.

\subsection{Traditional Approaches to Measuring Precision}

The precision of a static analyzer is usually judged by separating the \textit{true bugs} it found from the number of \textit{false alarms} it raised.\footnote{Note that in some cases, precision is measured through the proxy of another construction, such as generated call graphs \cite{DBLP:conf/popl/SmaragdakisBL11,helm2024total}.}
In practice, this measure requires a \textit{baseline} to be established.
This baseline requires tedious manual work discriminating alarms, which is almost impossible to realize on new analyses of large-scale projects.
We start by describing the case of manually annotated benchmarks where an absolute precision measure can thus be computed.
We then survey usual ways to measure precision when no baseline is known -- i.e, the number of \textit{true bugs} is not known.
\paragraph{Absolute precision: the case of manually annotated benchmarks.}
Some specific benchmarks have been manually crafted, or studied, to know where the \textit{true bugs} lie. This is for example the case of NIST's Juliet test suite for static analyzers by \cite{black2018juliet}, which contains tests labeled either as safe (no runtime errors) or buggy (with a single runtime error).
In that case, establishing an absolute precision measure is possible, provided of course that the classification is correct.
We leverage parts of the Juliet test suite to detect potential regressions of Mopsa in our continuous integration (cf. \Cref{sec:prec:ci}).
We show in \Cref{fig:juliet} our current precision results, measured as the percentage of tests where Mopsa is optimally precise.

\paragraph{When no baseline exists: usual approaches to measuring precision.}
In most cases however, manually checking all alarms to separate true alarms from the others is not possible for the purpose of experimental evaluations made in academia.
In those cases, other precision measures exist.
A first measure, which can be useful on small testcases, is recording a boolean holding when the program is safe.
This measure however is extremely coarse, which makes it unpractical to use during experimental evaluations or to discover improvements in an analysis.
A second approach is to report the number of alarms.
In our experience, this absolute number is however not informative: the numbers are difficult to put in context and might puzzle the community.
This can be mitigated by measuring number of alarms per total lines of code, but this measure will be highly application-dependent.

Note that other approaches to quantify precision can be leveraged when the goal is to compare the results of two analyses of the same program.
One can measure and compare abstract states at similar points of the analyses.
This measure however yields more questions than it solves: which abstract states should be compared?
For example, comparing all abstract states in the same contexts might be time-consuming due to the sheer number of states.
In addition, some precision changes will propagate from one state to its successor(s), which may skew the measure.
Additional issues can arise when incomparable domains are used.
In related work, the clam static analyzer from \cite{DBLP:conf/vstte/GurfinkelN21,DBLP:conf/cav/GurfinkelKKN15} provides a \texttt{clam-diff} utility to compare two different analyses of the same program, which relies on a semantic comparison of the numerical abstract states computed in each analysis.
This approach has more granularity than the one we currently use, and \cite{DBLP:conf/pldi/ArceriDZ23} have used it to evaluate precision changes in their work.
However, this approach may be too sensitive and it could be interesting to quantify the difference between some comparisons.
For example, the abstract state $a_1 = x < 1$ is included in $a_2 = x \leq 1$, which itself is included in $a_3 = \top$, but the change from $a_2$ to $a_3$ denotes a bigger precision loss.
To the best of our knowledge, this problem of quantifying the difference between two abstract states has only been considered by \cite{sotin:inria-00457324}.

\subsection{Reporting static analysis results in an automatic and transparent fashion}

\begin{figure}[!t]
  \begin{minipage}[t]{.21\textwidth}
\begin{minted}[escapeinside=||]{ocaml}
(* |$a^\#$| abstract state,
   |$p^\#$| safety property *)
if |$a^\# \not \sqsubseteq p^\#$| then
  add_alarm |$a^\#$| |$p^\#$|
\end{minted}
  \end{minipage}$\qquad\to\qquad$\begin{minipage}[t]{.18\textwidth}
\begin{minted}[escapeinside=||]{ocaml}
if |$a^\# \not \sqsubseteq p^\#$| then
  add_alarm |$a^\#$| |$p^\#$|
else
  add_safe_check |$p^\#$|
\end{minted}
  \end{minipage}
\caption{High-level implementation change, to move from reporting alarms to a transparent report of alarms and successful checks.}
\label{fig:implem:checks}
\end{figure}

\begin{figure}
  \includegraphics[width=.49\textwidth]{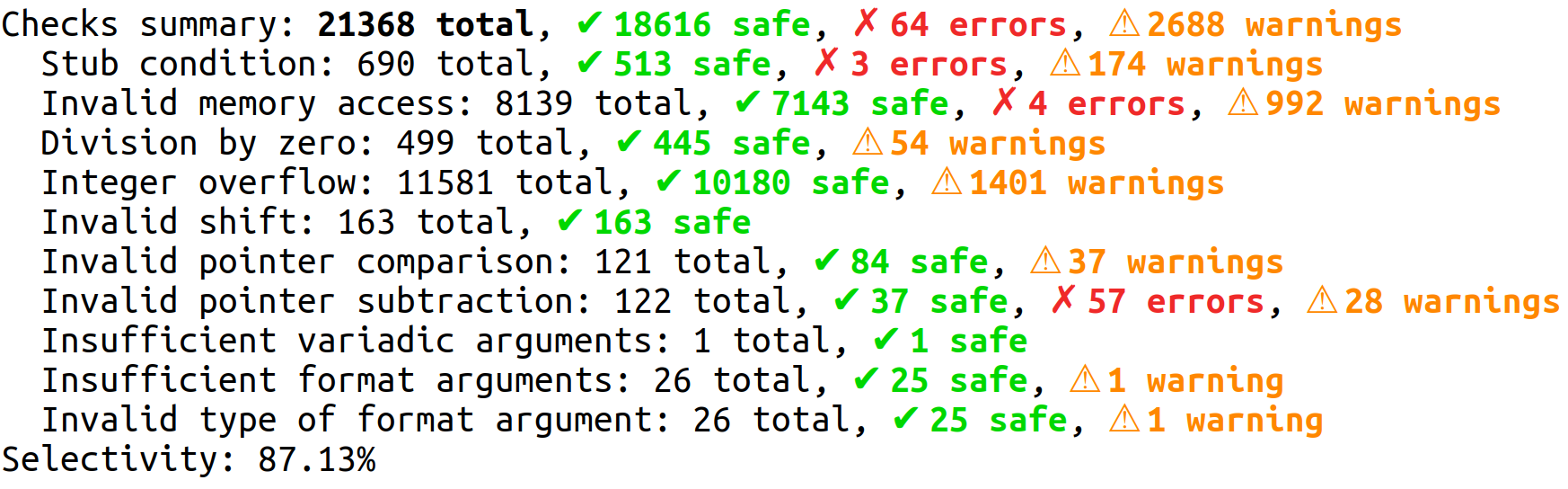}
  \caption{Analysis report summary for the analysis of \texttt{coreutils} \texttt{fmt}.}
  \label{fig:fmt}
\end{figure}

\begin{figure}
    \begin{subfigure}[T]{.4\textwidth}
\begin{minted}[escapeinside=||,numbers=left,xleftmargin=2em]{c}
int main(int argc, char** argv) {
  int y = -1;
  for(int x = 0; x < argc; x++) 
    y++;
}
\end{minted}
      \label{fig:selectivity:toy:code}
    \caption{Toy C example.}
    \end{subfigure}
    \begin{subfigure}[T]{.5\textwidth}
    \begin{tabular}{llll}
      \toprule
      \diagbox[innerwidth=2cm]{Stmt.}{Analysis} & {Intervals} & {Polyhedra}\\
      \midrule
      \texttt{x++}    & {Safe}      & {Safe}      \\
      \texttt{y++}    & {Alarm}     & {Safe}\\
      \midrule
      Selectivity     & {50\%}      & {100\%}\\
      \bottomrule
    \end{tabular}
    \caption{Selectivity measurement, in the case of integer overflow detection for the toy example of \Cref{fig:selectivity:toy:code}, analyzed either using intervals or a relational polyhedra abstract domain.}
  \end{subfigure}
  \caption{Illustrating selectivity computation on a toy C example.}
  \label{fig:selectivity:toy}
\end{figure}

We have developed an approach to report static analysis results in Mopsa which is both automatic and enhances the transparency of the verifications performed by the analysis.

Traditionally, static analyzers only report alarms, which correspond to failed proofs of safety: program locations\footnote{and callstacks, for our fully context-sensitive analyses.} where the abstract state does not satisfy a property the analyzer checks (such as absence of runtime errors).
Our approach consists in also logging successful proofs, which we name safe checks (of a given a property).
In the implementation, this change is conceptually easy to add, and boils down to the snippets shown in \Cref{fig:implem:checks}.
The analysis can then report all checks it performed, including alarms, and in a complementary way, the number of safe checks.
An example of real analysis report is shown in \Cref{fig:fmt}.
From this report, we can derive a numeric notion of precision we call \emph{selectivity}, measuring the ratio of successful proofs Mopsa has been able to perform.
In the case of the analysis of \texttt{coreutils} \texttt{fmt} in \Cref{fig:fmt}, the selectivity is 87\%.

\begin{equation*}
  \text{Selectivity} = \frac{\# \text{checks proved safe}}{\# \text{checks}}
\end{equation*}

We illustrate these notions of (safe) checks and selectivity computation on a toy example in \Cref{fig:selectivity:toy}.
In that case, we assume we are only interested in checking integer overflows, so we only check the two increments \texttt{x++} and \texttt{y++}.
In a configuration using a non-relational interval analysis, Mopsa will be able to prove that \texttt{x} stays in the range of signed integers, but will fail to prove it for \texttt{y}.
In that case, the selectivity is thus one half.
Moving to a relational analysis relying on polyhedra, Mopsa will be able to infer that \texttt{y} is bounded by \texttt{x}, and hence prove that both operations are safe from overflows, resulting in a selectivity of 100\%.

This approach has several benefits: it requires a lightweight implementation and provides transparent results, where users can clearly read in the analysis reports what the analysis has been able to verify.
Finally, selectivity provides a relative measure of precision, making it more informative than reporting an absolute number of alarms.
This measure depends on the complexity of the analyzed program expression and statements, rather than more arbitrary measures such as a program size.

When counting the number of checks (total and proved safe), it is possible to count the same location several times, as our analysis is fully context-sensitive: every context (i.e., call-stack) is counted separately. In general, comparing only the sensitivity of analyses with different coverage, causing the number of contexts to vary, can yield meaningless results. This justifies the introduction of fine-grained result comparison in the next subsection for those cases.
However, although Mopsa performs loop unrolling to improve precision, this does not affect the selectivity as unrolling is not part of the context: a check is considered safe if it is safe for all unrollings of the same context.

\subsection{Comparing analyses results using \texttt{mopsa-diff}}

\begin{figure*}
  \begin{subfigure}[T]{.55\textwidth}
\begin{minted}{text}
--- baseline/touch-many-symbolic-args-a4.json
+++ pplite/touch-many-symbolic-args-a4.json

- time: 589.0760
+ time: 675.1761

+ parse-datetime.y:1399.44-46: alarm: Invalid memory access
- parse-datetime.y:965.56-71: alarm: Invalid memory access
- parse-datetime.y:980.25-52: alarm: Invalid memory access
- parse-datetime.y:1003.23-50: alarm: Invalid memory access
- parse-datetime.y:921.56-71: alarm: Invalid memory access
- parse-datetime.c:1733.2-8: alarm: Invalid memory access
- parse-datetime.y:781.26-41: alarm: Invalid memory access
- parse-datetime.y:772.23-38: alarm: Invalid memory access
- parse-datetime.y:755.23-38: alarm: Invalid memory access
- parse-datetime.y:973.25-52: alarm: Invalid memory access
- parse-datetime.y:610.8-41: alarm: Invalid memory access
- parse-datetime.y:743.25-40: alarm: Invalid memory access
\end{minted}
    \caption{Comparing two analyses on a single program.}
    \label{diff:single}
  \end{subfigure}
 \hfill
  \begin{subfigure}[T]{.35\textwidth}
\begin{minted}{text}
139 reports compared
avg. time change       +52.065s
avg. speedup               -36%
new alarms                    2
removed alarms               32
new assumptions               0
removed assumptions           0
new successes                 0
new failures                  0
\end{minted}
    \caption{Comparing two analyses on all coreutils.}
\label{diff:multi}
  \end{subfigure}
  \caption{Mopsa-diff output comparing the impact of different relational domains on some \texttt{coreutils} programs.}
  \label{fig:diff}
\end{figure*}

\texttt{mopsa-diff} is a tool we developed which can compare analyses reports: it can either focus on comparing two analyses of a given program, or on comparing analyses results on a set of programs.
Comparing two analyses of a common program can be used to detect breaking changes (in terms of soundness or precision) when the implementation of the analyzer changes.
Similarly, we can compare the performance-precision benefits of changing configurations.
\texttt{mopsa-diff} can also be lifted to inspect the impact of a configuration change to a set of benchmarks for example.

These two modes are displayed in \Cref{fig:diff}, where we compare the analysis of some \texttt{coreutils} program with two relational numerical domains (chosen for illustrative purposes): PPLite v0.11 from \cite{pplite} and the NewPolka implementation of Apron v0.9.14 from \cite{apron}.
\Cref{diff:single} compares two analyses of \texttt{coreutils} \texttt{touch}.
The diff-like output shows that the PPLite version is able to remove 11 alarms compared to new Polka, although it is a bit slower, and adds another alarm.
\Cref{diff:multi} provides an overall comparison on all coreutils.
We can notice that a few alarms are removed by this version, although two new alarms are generated.
The analysis is overall slower.
No new soundness assumptions\footnote{During an analysis, Mopsa may make specific assumptions impacting the soundness of the analysis. Mopsa takes great care to report any such assumptions, and issues a special warning for the user to check them when there are some, for the sake of transparency. For example, Mopsa assumes by default that external C functions have no side effects on their parameters or global variables. We believe this approach to be a practical take on the principles highlighted in the soundiness manifesto of \cite{DBLP:journals/cacm/LivshitsSSLACGKMV15}.} are recorded, and the PPLite analysis does not crash on any program NewPolka is able to analyze.

\subsection{Detecting breaking changes during continuous integration}
\label{sec:prec:ci}

We leverage the analyses reports and \texttt{mopsa-diff} in our continuous integration to detect breaking changes affecting the soundness and precision of the analysis.
This is done by comparing obtained results with baseline results.
The set of benchmarks used in our continuous integration corresponds to all open-source projects we have analyzed in past experimental evaluations of our works.
We strive not to modify the source code in order to stick close to a real-world usage.
In some highly exceptional cases, we may rely on stubs to improve the results.
We currently have just one stub for all benchmarks used in our continuous integration.
\Cref{fig:juliet,fig:benches_loc,fig:benchs} show the benchmarks we currently use.
Running times have been measured on a desktop machine relying on an Intel Core i7-12700.

\begin{table}[!t]
  \centering
  \begin{tabular}{HlrrrrH}
    \toprule
                                                      & Benchmark &    \# Tests & Total LOC      & Time            & Precision & Imprecise\\
    \midrule
    \multirow{12}{*}{\rotatebox{90}{\textbf{Juliet}}}  &  CWE121  &    2,508    &  234,930       &      3,064s     &   22.13\%    &   77.87\%   \\
                                                       &  CWE122  &    1,556    &  166,664       &      1,948s     &   25.84\%    &   74.16\%   \\
                                                       &  CWE124  &     758     &   93,372       &      961s       &   36.94\%    &   63.06\%   \\
                                                       &  CWE126  &     600     &   75,984       &      769s       &   46.83\%    &   53.17\%   \\
                                                       &  CWE127  &     758     &   89,022       &      963s       &   37.07\%    &   62.93\%   \\
                                                       &  CWE190  &    3,420    &  440,749       &      4,356s     &   78.13\%    &   21.87\%   \\
                                                       &  CWE191  &    2,622    &  340,884       &      3,236s     &   78.87\%    &   21.13\%   \\
                                                       &  CWE369  &     497     &   83,238       &      674s       &   70.42\%    &   29.58\%   \\
                                                       &  CWE415  &     190     &   17,990       &      228s       &   100.00\%   &    0.00\%   \\
                                                       &  CWE416  &     118     &   14,782       &      142s       &   67.80\%    &   32.20\%   \\
                                                       &  CWE469  &     18      &    1,520       &       22s       &   100.00\%   &    0.00\%   \\
                                                       &  CWE476  &     216     &   20,427       &      254s       &   100.00\%   &    0.00\%   \\
    \bottomrule
  \end{tabular}
  \caption{Results of Mopsa analysis on Juliet benchmarks (non-relational configuration, no partitioning). CWE121 contains 2,508 tests, which overall take 3,064s to analyze. Mopsa is able to pass 22.13\% with a precise analysis, and is imprecise in the 77.87\% remaining cases.}
  \label{fig:juliet}
\end{table}
\begin{table}
  \centering
  \begin{tabular}{lrr}
    \toprule
    Benchmark     & LOC (C) & LOC (Python)\\
    \midrule
    \href{https://gitlab.com/mopsa/benchmarks/coreutils-benchmarks}{coreutils}     & 148,214 &          0 \\
    \href{https://gitlab.com/mopsa/benchmarks/pyperformance-benchmarks}{pyperformance} &       0 &      4,215  \\
    \href{https://gitlab.com/mopsa/benchmarks/pathpicker-analysis}{fpp}           &       0 &      3,140  \\
    \href{https://gitlab.com/mopsa/benchmarks/cpython-benchmarks/bitarray-analysis}{bitarray}      &   2,969 &      2,474  \\
    \href{https://gitlab.com/mopsa/benchmarks/cpython-benchmarks/cdistance-analysis}{cdistance}     &     912 &        979  \\
    \href{https://gitlab.com/mopsa/benchmarks/cpython-benchmarks/levenshtein-analysis}{levenshtein}   &   5,120 &        357  \\
    \href{https://gitlab.com/mopsa/benchmarks/cpython-benchmarks/llist-analysis}{llist}         &   2,757 &      1,686  \\
    \href{https://gitlab.com/mopsa/benchmarks/cpython-benchmarks/noise-analysis}{noise}         &     636 &        631  \\
    \href{https://gitlab.com/mopsa/benchmarks/cpython-benchmarks/pyahocorasick-analysis}{pyahocorasick} &   2,933 &      1,336  \\
    \bottomrule
  \end{tabular}
  \caption{Link and lines of code (loc) of each project Mopsa uses in its continuous integration. LOCs have been measured using cloc from \cite{cloc}.}
  \label{fig:benches_loc}
\end{table}

\begin{table}
  \centering
  \begin{tabular}{llrrr} \toprule
    \multicolumn{2}{c}{Benchmark}                                & Time & Selectivity & \# checks\\
    \midrule
\multirow{32}{*}{\rotatebox{90}{\textbf{coreutils}}} & basename                & 33.79s           & 98.65\%        & 11,731           \\
                                                     & comm                    & 42.67s           & 97.32\%        & 12,654           \\
                                                     & dircolors               & 34.82s           & 99.74\%        & 20,062           \\
                                                     & dirname                 & 21.68s           & 99.61\%        & 11,307           \\
                                                     & echo                    & 19.26s           & 99.43\%        & 11,010           \\
                                                     & false                   & 14.50s           & 99.72\%        & 10,774           \\
                                                     & getlimits               & 34.62s           & 98.54\%        & 11,711           \\
                                                     & hostid                  & 18.05s           & 99.65\%        & 11,303           \\
                                                     & id                      & 32.69s           & 99.04\%        & 12,338           \\
                                                     & link                    & 23.03s           & 99.52\%        & 11,572           \\
                                                     & logname                 & 20.36s           & 99.66\%        & 11,307           \\
                                                     & mkfifo                  & 34.87s           & 99.20\%        & 11,807           \\
                                                     & mknod                   & 34.98s           & 99.11\%        & 12,513           \\
                                                     & nice                    & 23.36s           & 99.55\%        & 11,463           \\
                                                     & nohup                   & 26.98s           & 99.27\%        & 11,734           \\
                                                     & nproc                   & 17.37s           & 99.44\%        & 11,533           \\
                                                     & printenv                & 23.59s           & 99.50\%        & 11,202           \\
                                                     & pwd                     & 22.04s           & 99.62\%        & 11,502           \\
                                                     & rmdir                   & 39.00s           & 99.22\%        & 11,699           \\
                                                     & runcon                  & 18.55s           & 99.66\%        & 11,215           \\
                                                     & seq                     & 42.68s           & 95.87\%        & 14,310           \\
                                                     & sleep                   & 23.79s           & 99.46\%        & 11,546           \\
                                                     & stdbuf                  & 32.16s           & 98.46\%        & 12,526           \\
                                                     & sync                    & 24.53s           & 99.60\%        & 11,273           \\
                                                     & tee                     & 35.69s           & 98.76\%        & 12,057           \\
                                                     & timeout                 & 32.28s           & 98.51\%        & 12,420           \\
                                                     & true                    & 9.55s            & 99.72\%        & 10,774           \\
                                                     & uname                   & 20.61s           & 99.52\%        & 11,943           \\
                                                     & unlink                  & 16.17s           & 99.63\%        & 11,497           \\
                                                     & users                   & 20.82s           & 99.06\%        & 11,668           \\
                                                     & whoami                  & 13.03s           & 99.66\%        & 11,329           \\
                                                     & yes                     & 19.82s           & 99.45\%        & 11,216           \\
    \midrule
\multirow{12}{*}{\rotatebox{90}{\textbf{pyperformance}}}    & chaos                & 8.55s            & 98.86\%        & 6,415            \\
                                                     & fannkuch             & 0.31s            & 98.75\%        & 321              \\
                                                     & float                & 0.10s            & 99.48\%        & 574              \\
                                                     & go                   & 143.72s          & 97.67\%        & 7,552            \\
                                                     & hexiom               & 39.30s           & 98.33\%        & 5,392            \\
                                                     & nbody                & 1.28s            & 99.73\%        & 1,100            \\
                                                     & raytrace             & 49.99s           & 98.95\%        & 5,694            \\
                                                     & regex\_v8             & 18.30s           & 99.56\%        & 32,638           \\
                                                     & richards             & 11.86s           & 99.65\%        & 3,710            \\
                                                     & scimark              & 22.74s           & 99.09\%        & 6,397            \\
                                                     & spectral\_norm        & 1.30s            & 99.43\%        & 697              \\
                                                     & unpack\_sequence      & 3.46s            & 100.00\%       & 8,094            \\
    \midrule
\multirow{2}{*}{\rotatebox{90}{{\textbf{fpp}}}}& choose              & 155.27s          & 99.76\%        & 91,417           \\
                                                     & processInput        & 3.84s            & 99.78\%        & 4,914            \\
    \midrule
                                                    & bitarray           & 499.26s          & 89.57\%        & 288,475          \\
    \midrule
                                                    & cdistance            & 58.26s           & 96.54\%        & 46,512           \\
    \midrule
                                                    & levenshtein   & 27.26s           & 85.35\%        & 15,519           \\
    \midrule
                                                    & llist              & 78.70s           & 98.92\%        & 136,237          \\
    \midrule
\multirow{2}{*}{\rotatebox{90}{\textbf{noise}\hspace{-.1em}}}     & perlin       & 5.80s            & 99.20\%        & 4,273            \\
                                                    & simplex      & 6.06s            & 99.45\%        & 4,586            \\
    \midrule
                                                    & pyahocorasick & 29.92s           & 89.67\%        & 20,415           \\
    \bottomrule
  \end{tabular}
  \caption{Benchmarks used in Mopsa's non-regression testsuite. \texttt{coreutils} benchmarks have been analyzed with fully symbolic arguments, and a relational analysis. Other projects have been analyzed in a non-relational setting.}
  \label{fig:benchs}
\end{table}

\section{Instrumenting the Analysis}
\label{sec:hooks}

Hooks are plug-ins that can observe, and influence, the analysis.
When enabled, these hooks are called before and after every statement is analyzed, and they can peek at the input and output abstract states.
Hooks do not have access to the private, internal representation of abstract domains, but communicate through a public interface of queries (described by \cite{mopsa}).

These hooks have a wide variety of usages ranging from providing interpretation traces to improving the precision by detecting relevant thresholds which can be used by the widening.
In the context of debugging and maintaining a static analyzer, these hooks are handy because they work at the level of the \emph{abstract execution of the input program}, while standard profiling and coverage tools can provide information about the \emph{execution of the static analyzer}.
In the remainder of this section, we showcase computations of abstract coverage, abstract profiling, and heuristic detection of unsoundness and large imprecision.

\subsection{Coverage}

This hook computes the abstract coverage of statements that have been reached by the analysis.
It gives a birds-eye view of where the analysis went, and can easily be leveraged to detect insufficient modeling assumptions reducing the search space of the analysis, as the user can quickly spot functions that should be reachable but are not.
These modeling assumptions can be due to the modeling of command-line arguments.

As an example, the analysis of \texttt{coreutils} \texttt{fmt} shows that 76\% of its \texttt{main} function is covered, when the analysis simply assumes that no arguments are passed to the utility.
Given that \texttt{fmt} is a command-line utility, assuming that no arguments are passed is too restrictive.
We can thus move to an analysis which performs a symbolic modeling of arguments: it considers that the analyzed program arguments are an array of arbitrary size (within either system bounds or user-supplied bounds) containing strings of arbitrary size and contents covering all actual possible usage.
In this case of symbolic modeling of arguments, we reach 100\% coverage for the \texttt{main} function.
Note that by default, the analysis makes no restriction on the arguments of the \texttt{main} entry point (i.e, arguments are modeled symbolically by default).

This hook can help users find soundness issues related to their configuration and instrumention of the analysis.

\subsection{Profiling}

\begin{figure*}[!t]
  \includegraphics[width=\textwidth]{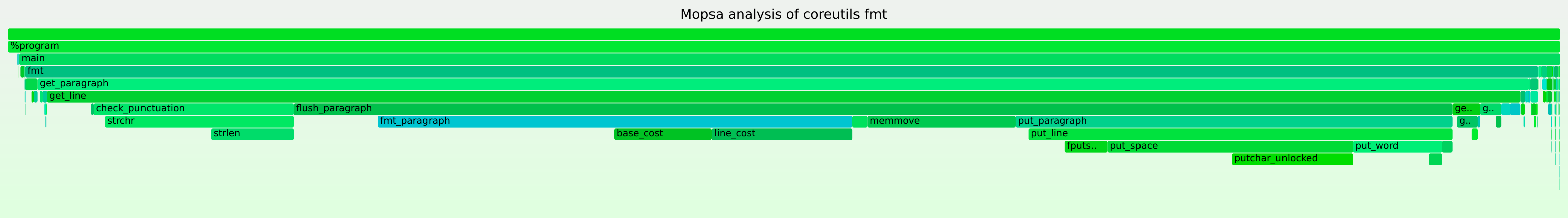}
  \caption{Flamegraph obtained using the abstract function profiling of Mopsa, when analyzing \texttt{coreutils} \texttt{fmt}.}
  \label{fig:flamegraph}
\end{figure*}

Loops and function calls are the two reasons why the same block of code might need to be analyzed a large number of times (due to loop iterations in one case, and context-sensitivity in the other case).
They are thus the main source of analysis cost, and it is important to be able to pinpoint which ones might be problematic in a specific analysis.
Mopsa thus provides two profiling hooks: one for loops and one for function calls.

The loop hook tracks the number of times a given loop is called, the number of iterations needed to reach a fixpoint, as well as the total time spent analyzing each loop.
This is helpful to fine-tune the widening, to track slowdowns in the postfixpoint computation performed by Mopsa, and more extremely, non-termination issues.

The function profiling hook tracks the number of times a function is analyzed as well as the time spent analyzing it.
This hook is particularly important as the current analyses written in Mopsa are fully context-sensitive, meaning they analyze functions by virtual inlining.
The hook thus helps identifying which functions take the most time to analyze.
In particular, the collected data can be transformed into the flamegraph graphical representation from \cite{DBLP:journals/cacm/Gregg16}; an example is shown in \Cref{fig:flamegraph}.
Tracking the number of times the function is analyzed is also important, as a frequent reason a function is reported as taking a large time is when the function is called many times, within nested loops, even if each function analysis is actually fast.

Hooks can produce and return information during abstract computations, without having to wait for the analysis to terminate.
In particular, they work on partial executions of the analysis, which is especially useful for
profiling programs where the analysis is unexpectedly long.
It will only give a partial picture but will highlight the relevant loops and functions taking time to be analyzed.

\subsection{Heuristic Unsoundness/Imprecision Detection}
\label{sec:heuristic:unsoundness}

We developed plugins performing heuristic detection of unsound or highly imprecise behaviors, which are reported to the user.
Most users will not precisely know the behavior of all domains an analysis configuration uses, especially due to the highly distributed nature of transfer functions in Mopsa.
It is thus interesting to have hooks acting as runtime mechanisms that can warn the user of unsound or highly imprecise behavior, making it easier to pinpoint a source of imprecision or unsoundness.

The unsoundness detector acts as a sanity check.
It currently verifies the following property: an assignment from a non-bottom state cannot return bottom\footnote{Except if the expression contains a runtime error, such as a division by zero.}.
Other properties could easily be added to this detector in future work.
While the coverage hook finds modeling and configuration errors from users, this detector finds soundness issues due to bugs in the analyzer.
These errors are critical, and corrected as soon as possible, as the analyses are constructed to be sound. 

The imprecision detector warns when the analysis of a (sub)expression yields top, which is the source of large imprecision.
Precision issues are less critical in Mopsa, as they usually result in a performance-precision tradeoff. 

\section{Abstract Debugging through an Interactive Engine}
\label{sec:interactive}

Traditional approaches to debugging static analyzers do not scale well.
We briefly survey basic techniques -- all supported by Mopsa -- and their shortcomings.
A first approach is to check the analysis output. It is usually quite coarse, and when unexpected behaviors happen, the output is not sufficient to get explanations.
Another approach is to provide some builtin functions asking for the abstract interpreter to print its current state.
Those can then be added in the source program to show specific abstract states.
\cite{kosmatov2024guide} mention the use of \texttt{Frama\_C\_show\_each} in Frama-C; Mopsa provides a similar \texttt{mopsa\_print}.
While this approach helps making the result understandable, the cost is somewhat high: the modification of the source code requires to restart the analysis each time, which can be prohibitive when programs are large.
If the program location where the printing happens is reached in a lot of different contexts, the output may turn out to be too verbose to be useful.
Additionally, this restart of the analysis can become more complicated if the precise location of the origin of the error is different from the location where it becomes manifest.
In that case, it must be discovered by trial and error, by inserting logging commands and running the analysis again many times.
A complementary approach is to store relevant analysis states and information so that users can inspect them afterwards.
Thus, a single analysis run can produce a log that can be examined at different program locations, even if these locations are not known in advance.
The Goblint tool from \cite{goblint-soft} provides an interactive output allowing to explore various abstract states after the analysis is finished, either as an HTML page or through the Debug Adapter Protocol available in different IDEs and pioneered by VSCode.
In Mopsa, we can record an interpretation trace, showing the order in which expressions and statements are analyzed, optionally with the abstract state.
This however can quickly become too big to process.
For example, the interpretation trace of analyzing \texttt{coreutils} \texttt{fmt} in Mopsa is around 12GB of text.

\begin{figure*}[!ht]
\centering
  \begin{subfigure}[T]{.5\textwidth}
      \includegraphics[width=\textwidth]{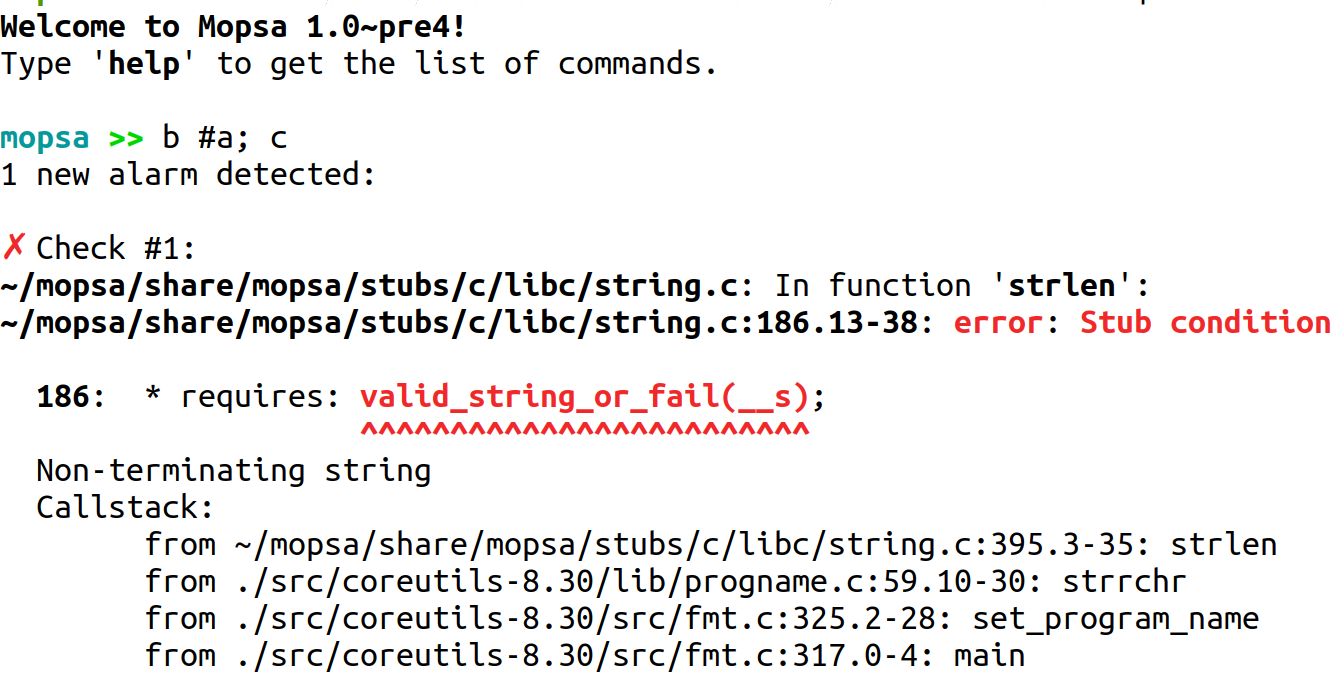}
      \caption{Starting interface of interactive engine, with first command typed, to stop the analysis at the first alarm.}
      \label{interactive:int:alarm}
    \end{subfigure}
\begin{minipage}[t]{\textwidth}
    \vspace{0pt}
    \begin{subfigure}[T]{.47\textwidth}
      \includegraphics[width=\textwidth]{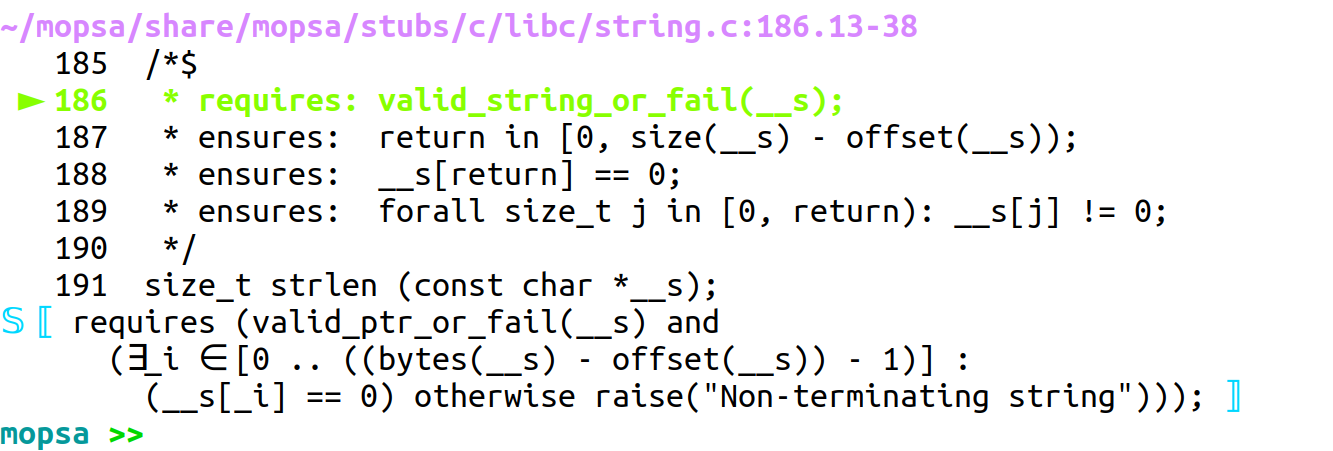}
      \caption{The engine jumped back to the beginning of the statement generating the alarm.}
      \label{interactive:int:prompt}
    \end{subfigure}
    \hfill
    \begin{subfigure}[T]{.5\textwidth}
      \includegraphics[width=\textwidth]{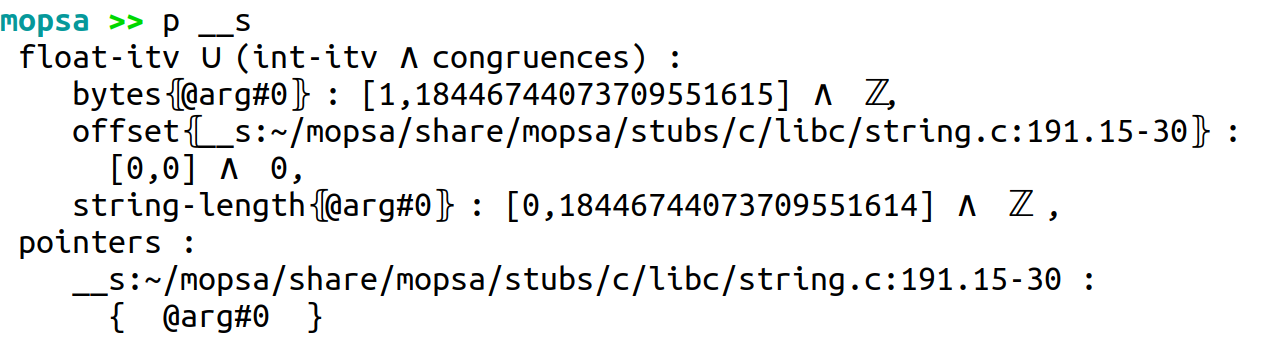}
      \caption{Interval analysis: Inspecting abstract information related to \texttt{\_\_s}.}
      \label{interactive:int:print}
    \end{subfigure}
    \begin{subfigure}[T]{.47\textwidth}
      \includegraphics[width=\textwidth]{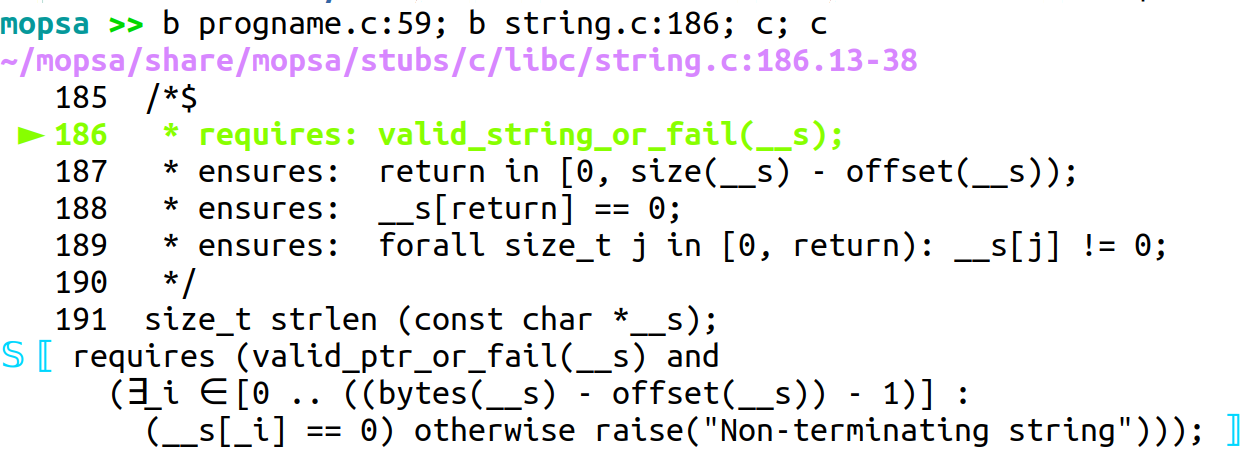}
      \caption{Relational analysis started, with breakpoints to reach the alarm detected in \Cref{interactive:int:alarm}.}
      \label{interactive:rel:prompt}
    \end{subfigure}
    \hfill
    \begin{subfigure}[T]{.5\textwidth}
      \includegraphics[width=\textwidth]{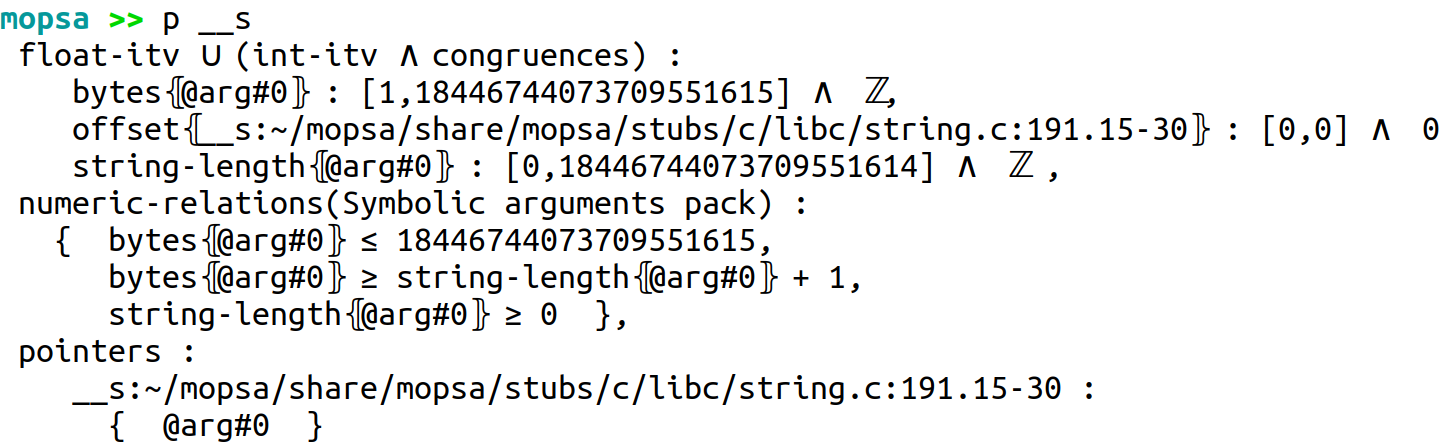}
      \caption{Relational analysis: Inspecting abstract information related to \texttt{\_\_s}.}
      \label{interactive:rel:print}
    \end{subfigure}
  \end{minipage}
  \caption{Interactive engine workflow example on the analysis of \texttt{coreutils} \texttt{fmt}.}
  \label{fig:interactive}
\end{figure*}

In this section, we showcase an abstract debugger, providing an interactive interface to the user.
The user decides how they want to navigate the abstract execution of the program, and computations are performed on-the-fly accordingly.
This debugger provides an interface similar to \texttt{gdb}, except that it works on the abstract execution of the program while \texttt{gdb} would work on a concrete execution.
As Mopsa currently works as an interpreter on the AST (and not as a generic equation solver), the flow of execution of the analysis is close to that of the concrete execution of the program, and easy to understand.
The abstract execution can be navigated using diverse strategies, such as going to the next statement, entering inner analysis of functions, or continuing until the next breakpoint.
It is also possible to observe intra-instruction analysis, and rewriting operations that are at the heart of Mopsa.
Breakpoints can be program locations, functions, transfer functions or the next detected alarm.
The abstract state can be printed, as well as projections of abstract information related to selected abstract variables.

Combined with a terminal multiplexer such as \texttt{tmux}, this interactive interface can be used to perform some side-by-side debugging of the analysis of the same program in two different configurations.
Thanks to some wrapper scripts we have developed, commands for the interactive engine have to be entered once and will be given to both sides of the analysis, further easing debugging.

We show how the interactive engine can be used in \Cref{fig:interactive}, on an example where we analyze \texttt{coreutils} \texttt{fmt}. The analysis is fully context-sensitive, and supposes symbolic arguments are passed to \texttt{fmt}.

The interactive engine is shown just after it was started in \Cref{interactive:int:alarm}. The prompt (represented by \texttt{mopsa >>}) has been given two commands: \texttt{breakpoint \#alarm} and \texttt{continue} (abbreviated as \texttt{b \#a} and \texttt{c}, and chained using a semicolon). This adds a breakpoint at the next encountered alarm and runs the analysis until this breakpoint is reached. The analysis raises an alarm in a call to builtin function \texttt{strrchr} at \texttt{progname.c:59}.

Once the first alarm is reached, the engine jumps back to the state and statement reached just before the alarm is triggered, easing inspection and understanding of the issue. Thanks to the functional implementation of Mopsa, this jumping back in time is easy to implement.\footnote{We are experimenting with ways to enable going further backwards in the analysis, which would provide the interface of a reverse debugger.}
The state is shown in \Cref{interactive:int:prompt}.
The analysis currently checks that \texttt{strlen} can be called, through some preconditions encoded in the stub contract language of \cite{mopsa-cstubs}.
The targeted precondition (at line 186) aims at verifying that the string passed to \texttt{strlen} is valid.
The analysis is unable to prove the string is correctly encoded with a \texttt{'$\backslash$0'} at its end, as expressed by the existentially quantified formula.

We can take a look at the abstract state to understand why the formula above cannot be proved correct (\Cref{interactive:int:print}).
There is no need to read the full abstract state, we can just ask for relevant abstract information related to \texttt{\_\_s} with \texttt{print \_\_s} (\texttt{print} can be abbreviated by \texttt{p}). We learn it points to a memory block of undetermined size ($\texttt{bytes}(\texttt{@arg\#0})$), and the position of the first 0 in the string (i.e, the auxiliary variable representing string lengths, $\texttt{string-length}(\texttt{@arg\#0})$, as defined by \cite{DBLP:conf/sas/JournaultMO18}) is not known precisely either. Due to the non-relationality of the interval abstract domain, the analysis is thus unable to prove that the string \texttt{\_\_s} has a terminating character. 

Let us restart the analysis in a relational setting, relying on the polyhedra abstract domain. In order to scale we use static packing (introduced by \cite{astree}), a technique that keeps multiple polyhedra of small dimensions rather than a high-dimension polyhedra. We introduce here a specific pack handling the symbolic arguments used to analyze the program. In \Cref{interactive:rel:prompt}, we add breakpoints to reach the program location where the alarm was raised with the interval analysis (\Cref{interactive:int:alarm}).

If we break to the program position where the alarm was raised with the interval analysis, we notice the numeric relations of \Cref{interactive:rel:print} express exactly what is needed to prove the existentially quantified formula from \Cref{interactive:int:prompt}. This change of configuration means the first alarm has been removed by moving to a more expressive domain.

\begin{figure*}
  \begin{subfigure}[T]{.475\textwidth}
    \includegraphics[width=1.05\textwidth]{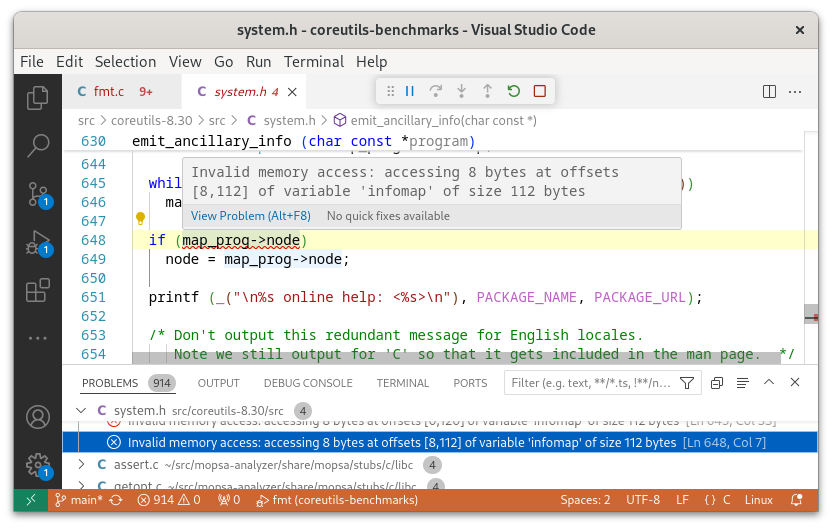}
    \caption{The center panel shows source code annotated with one of the alarms reported by Mopsa through the LSP, and the list of alarms in the bottom panel.}
    \label{vscode:lsp}
  \end{subfigure}
  \hspace{1em}
  \begin{subfigure}[T]{.475\textwidth}
    \includegraphics[width=1.05\textwidth]{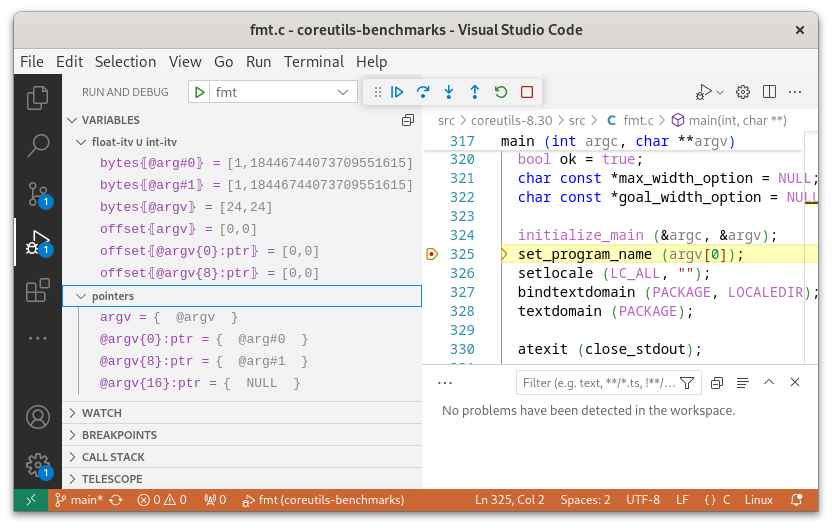}
    \caption{Using the DAP, a user can set a breakpoint at line 325 (right panel), and then inspect the program state (left panel).}
    \label{vscode:dap}
  \end{subfigure}
  \caption{Analysis of \texttt{coreutils} \texttt{fmt} through VSCode's interface.}
  \label{vscode}
\end{figure*}

\paragraph{From command-line interfaces to IDEs.}
The whole interface of Mopsa, from its batch mode providing alarm reports to its abstract debugger, are available through the command-line.
We provide similar interfaces for IDE users, by leveraging the Language Server Protocol (LSP) and the Debug Adapter Protocol (DAP) originally developed by VSCode.
In particular, we can report results of our analysis directly on the source code thanks to the LSP \Cref{vscode:lsp}, just as linters do.
We use the DAP to provide interactive, abstract debugger sessions within the IDE, highlighted in \Cref{vscode:dap}.

\section{Automated Testcase Reduction}
\label{sec:creduce}

Testcase reduction is an automated approach aiming at minimizing a test while keeping a specific property.
One tool performing automated testcase reduction for C is \texttt{creduce}, developed by \cite{creduce}.
It has originally been applied to compilers (where a test is an input program), and enjoys widespread use in this case.
For example, \cite{gcc-creduce} provide a guide to testcase reduction and asks for reduced testcases in their bug reports.
We report our successful use of \texttt{creduce} on static analyzers, which are similarly highly complex pieces of code, manipulating potentially large input programs.

The \texttt{creduce} workflow is summarized in \Cref{fig:creduce}.
The user provides an input testcase (\texttt{file.c}) exhibiting an unwanted behavior (such as a crash).
An oracle, written as a shell script, describes whether the unwanted behavior still occurs in partially reduced versions.
\texttt{creduce} will then loop until it reaches a minimal testcase satisfying the oracle.
One challenge of putting automated testcase reduction into practice is to establish sufficiently robust, yet automated, testcase oracles.

We start by describing two ways to leverage automated testcase reduction to ease debugging in Mopsa in \Cref{sec:sub:red}.
Our first usecase concerns internal errors from the analyzer that can happen deep into long analyses.
Our second usecase pinpoints soundness issues by comparing two analyses, yielding different results on the same program.
We finish by highlighting in \Cref{sec:mopsa2creduce} that the collaboration between Mopsa and \texttt{creduce} is a two-way street, as Mopsa can simplify testcase reduction for multi-file projects with complex commands such as \texttt{GNU coreutils}.

\begin{figure}
  \tikzstyle{Box2}=[rectangle, draw, rounded corners, fill= white, align= center, copy shadow={draw, shadow xshift=0.5mm, shadow yshift=-0.5mm}]
  \tikzstyle{Box}=[draw, text centered]
  \centering
  \begin{tikzpicture}[scale=1,font=\small]

    \node (cprog) [Box] {\texttt{file.c}};
    \node (oracle) [Box, below = of cprog] {\texttt{oracle.sh}};

    \node (creduce) [Box, below right = 0.25 and 2em of cprog] {\texttt{creduce}};
    \node (reduced) [Box,rounded corners, right = 1.5em of creduce] {\texttt{small.c}};

    \draw[-stealth,line width=1pt] (cprog) -- (creduce);
    \draw[-stealth,line width=1pt] (oracle) -- (creduce);
    \draw[-stealth,line width=1pt] (creduce) -- (reduced);
  \end{tikzpicture}
  \caption{creduce pipeline.}
  \label{fig:creduce}
\end{figure}

\subsection{Leveraging Automated Testcase Reduction}
\label{sec:sub:red}

\paragraph{Internal error reduction.} We have successfully used \texttt{creduce} to pinpoint and fix internal errors within Mopsa which result in uncaught OCaml exceptions terminating the analysis.
In that case, the oracle we use first checks that the produced file can be compiled without errors\footnote{\texttt{creduce} can produce C programs with illegal syntax or types, that are naturally rejected (early) by a compiler's frontend.}, and then checks that the same internal error is raised during the analysis of the reduced testcase.
In our experience, this kind of reduction only requires the exception name and message to be passed to the oracle.
We provide in \Cref{table:creduce:single} examples of issues or merge requests where automated testcase reduction has been used.
These cases stem from benchmarks from the Software-Verification Competition, or from \texttt{coreutils} programs.
Lines of code correspond to the number of lines of preprocessed C code (cf. \Cref{sec:mopsa2creduce}), formatted using \texttt{clang-format}, and measured through command \texttt{wc -l}.
In our experience, reduction time is less than 12 hours, meaning that it can easily be run overnight and save a considerable amount of human time.
It helped us solve issues that were otherwise out of reach.
For example, issue 81 from \Cref{table:creduce:single} was reduced from one linux device driver code from the Software Verification Competition, and caused by an implementation bug in the string length domain from \cite{DBLP:conf/sas/JournaultMO18}.

\begin{table}
  \centering
  \begin{tabular}{rHrrr}
    \toprule
    Reference & Origin & Original LoC & Reduced LoC & Reduction\\
    \midrule
  \href{https://gitlab.com/mopsa/mopsa-analyzer/-/issues/76}{Issue 76} & SV-Comp & 28,737 & 18 & 99.94\% \\
  \href{https://gitlab.com/mopsa/mopsa-analyzer/-/issues/81}{Issue 81} & SV-Comp & 15,627 & 8 & 99.95\% \\
  \href{https://gitlab.com/mopsa/mopsa-analyzer/-/issues/134}{Issue 134} & SV-Comp & 17,411 & 10 & 99.94\% \\
  \href{https://gitlab.com/mopsa/mopsa-analyzer/-/issues/135}{Issue 135} & SV-Comp & 7,016 & 12 & 99.83\% \\
  \href{https://gitlab.com/mopsa/mopsa-analyzer/-/merge_requests/130#note_1516013076}{M.R. 130} & \texttt{coreutils} & 77,981 & 20 & 99.97\% \\
    \href{https://gitlab.com/mopsa/mopsa-analyzer/-/commit/34baaa483725cb81bacf6cc8144fc9c86a8bdd63}{M.R. 145} & \texttt{coreutils} & 77,427 & 19 & 99.98\% \\
    \bottomrule
  \end{tabular}
  \caption{Internal error reduction examples, taken from \href{https://gitlab.com/mopsa/mopsa-analyzer/}{the Gitlab repository of Mopsa}.}
  \label{table:creduce:single}
\end{table}

\paragraph{Differential-configuration reduction.}
\label{creduce:conf}

The testcase reduction for internal errors is the canonical usecase of such tools.
We have had some recent successes in \textit{differential-configuration reduction}, easing the debugging of cases where two different configurations of Mopsa (\Cref{sec:sub:precision}) yield contradictory analysis results on a given program.
We have typically applied it when an analysis is unsound, and where the culprit (abstract domain or reduction) is included in one configuration and not the other.
This is particularly useful given the large number of combinations of abstract domains Mopsa enables through its modular design.
We have used this approach to simplify some soundness issues reported by external users (\href{https://gitlab.com/mopsa/mopsa-analyzer/-/issues/179}{\# 179}, \href{https://gitlab.com/mopsa/mopsa-analyzer/-/issues/182}{\# 182}), who then integrated it in their process when reporting further issues (\href{https://gitlab.com/mopsa/mopsa-analyzer/-/issues/184}{\# 184}, \href{https://gitlab.com/mopsa/mopsa-analyzer/-/issues/185}{\# 185}).
Thanks to this approach, we have minimal programs in which to debug the source of unsoundness.
For example, the reduction from issue 182 allowed to quickly identify the division from the integer powerset abstraction as the source of unsoundness which was then easily fixed.
We have currently not explored how \texttt{creduce} could help us pinpoint precision improvements.

\subsection{Leveraging Mopsa to Ease Multi-file Reduction}
\label{sec:mopsa2creduce}

One of the current usability barriers to automated testcase reduction through \texttt{creduce} and its sibling \texttt{cvise} (developed by \cite{cvise}) is the support of multi-file projects.
Indeed, \texttt{creduce} requires the explicit list of files to be reduced\footnote{\url{https://github.com/csmith-project/creduce/blob/31e855e290970cba0286e5032971509c0e7c0a80/creduce/creduce.in\#L197}}.
This list can be difficult to establish on large open-source projects, such as coreutils, where a build system like \texttt{make} takes care of compiling the various sources into an executable, through a list of complex rules.
In addition, some large projects may use different files with different compilation options, which would create an additional difficulty in using standard \texttt{creduce}.

Mopsa natively supports the analysis of multi-file C projects, through a utility called \texttt{mopsa-build} which creates a compilation database by instrumenting the compilation process.
This process can include \texttt{configure} scripts, \texttt{make}, \texttt{cmake}, etc.
\texttt{mopsa-build} overrides environment variables to record all compiler and linker calls, as well as the options that were passed to them.
Due to its seamless nature, \texttt{mopsa-build} can be used as a drop-in replacement of various build systems such as \texttt{make}: \texttt{mopsa-build make}.
Then, this compilation database can be leveraged by the C analysis to analyze a specific target of the build system.
An important side-effect of this process is the ability of the analysis to generate a single preprocessed file\footnote{The single file is generated as a kind of source-level (or AST-level) linking.}, which does heavily simplify automated testcase reduction.

\section{Related Work}

To the best of our knowledge \cite{DBLP:conf/pldi/AndreasenMN17} are the first to describe approaches they use to improve their static analysis of JavaScript. They describe combinations of techniques relying on delta-debugging, soundness testing (through comparison with concrete values) and blended analysis (injection of concrete values to restrict the abstract state).
The automated testcase reduction we use is close to the delta-debugging techniques they rely on, but otherwise our approaches seem complementary.

\paragraph{Static analyzer interfaces.}
Our interactive engine can act as an abstract debugger, either from the command-line interface, or through IDEs supporting the debug-adapter protocol (DAP).
Some other analyzers can report their alarms through the language-server protocol (LSP) developed initially for linters.
A specific interface called MagpieBridge has been developed by \cite{DBLP:conf/ecoop/LuoDB19} to simplify the integration of static analyzers within LSP.
\cite{DBLP:conf/sle/MolleVR23} showcase a cross-level debugger, working on the analyzed program and enabling conditional breakpoints expressed either on the analyzed program or the state of the analyzer itself.
The Goblint static analyzer provides a different kind of abstract debugger developed by \cite{abs-debug}.
It provides an interactive, graphical exploration of the abstract states in the control-flow graph of the program once the analysis has successfully finished.
Both abstract debuggers from Goblint and Mopsa provide an IDE integration through the Debug Adapter Protocol.
In their documentation of the Goblint analyzer, \cite{goblint-doc} also mention automated testcase reduction, in particular to debug fixpoint termination issues.

A blog post from the Frama-C team by \cite{framac-creduce} highlights that automated testcase reduction simplified their interaction with industrial clients having private codebases.
At least for runtime errors, these industrial clients can run automated testcase reduction by themselves.
The reduction will yield a highly simplified testcase that will not leak important parts of the initial, private codebase, which can then be sent to Frama-C developers without any confidentiality issues.

\paragraph{Testing the soundness and precision of static analyzers.}
\cite{DBLP:conf/kbse/BugariuWC018} perform automated testing of numerical abstract domains, by checking that some chosen properties should be verified.
We rely on a similar yet simplified approach in \Cref{sec:heuristic:unsoundness}, through an encoding of a heuristic rule to detect unsoundness during an analysis.
However, our approach is not specific to numerical abstract domains.
Goblint relies on a similar approach implemented by \cite{goblint-issue} to check that at least one branch in conditionals is not dead.

\cite{DBLP:conf/issta/KlingerCW19} describe ways to automatically compare the soundness and precision of different C static analyzers on programs from the Software-Verification Competition (SV-Comp).
In particular, the original program can be mutated to check the results at different program points, and a notion of $\delta-$unsoundness is established, to reduce spurious warnings.
In our case, we have only considered different configurations of Mopsa, but not compared it using testcase reduction to other static analysis tools.
\cite{DBLP:conf/cgo/TanejaLR20} develop SMT-based algorithms that can detect soundness and precision errors of some of LLVM dataflow analyses.

Formally verified static analyzers, such as the work of \cite{DBLP:conf/popl/JourdanLBLP15}, will not require debugging of unsound results by construction.
However, we believe the techniques we presented could still be interesting to investigate precision issues.

\section{Conclusion}
This article documents and shares the practices we have established for the maintenance of the Mopsa static analyzer during the last 7 years.
In particular, we rely on a measure of precision than can be computed automatically on software without baselines on the number of true bugs.
This approach increases the transparency of the analysis, and simplifies regression detection.
We have shown different tools (profiler, debugger) focusing on the abstract execution of the analyzed program, and reported use of automated testcase reduction to simplify our debugging.
Following the work of \cite{DBLP:conf/pldi/AndreasenMN17}, we hope this article will inspire other groups and encourage other researchers to document and share their practices.

There are still some challenges revolving around the development of Mopsa.
The number of different configurations to analyze a given language can grow quite quickly due to the modular architecture of Mopsa.
We are currently performing regression tests on selected, specific configurations to reduce the computational cost.
Code maintenance and debugging can still take some time, and onboarding material takes a lot of time to create and maintain.
Finally, we are looking into providing an install-free version of Mopsa -- through a web page for example -- meaning that prospective users can quickly test it without any installation.

\bibliographystyle{spbasic}
\bibliography{paper}

\doclicenseThis

\end{document}